\documentclass[letterpaper, 10 pt, conference]{ieeeconf}  

\IEEEoverridecommandlockouts                              

\makeatletter
\let\NAT@parse\undefined
\makeatother
\usepackage{hyperref}
\usepackage{cite}
\usepackage{amsmath,amssymb,amsfonts}
\usepackage{algorithmic}
\usepackage{graphicx}
\usepackage{textcomp}
\usepackage{xcolor}
\usepackage{tabularx}
\usepackage{hyperref}
\hypersetup{
    colorlinks=true, 
    linkcolor=black,   
    citecolor=black,   
    urlcolor=black,    
}
\usepackage[utf8]{inputenc}
\usepackage[T1]{fontenc}
\usepackage{multirow}
\usepackage{pifont}
\usepackage{booktabs}
\usepackage{array}
\usepackage{makecell}
\usepackage{graphics} 
\usepackage{epsfig} 
\usepackage{mathptmx} 
\usepackage{times} 
\usepackage{amsmath} 
\usepackage{amssymb}  
\usepackage{array}

\title{\LARGE \bf
VNet: A GAN-based Multi-Tier Discriminator Network for Speech Synthesis Vocoders*
}

\author{Yubing Cao$^{1}$, Yongming Li$^{1\dagger}$, Liejun Wang$^{1}$ and Yinfeng Yu$^{1}$
\thanks{$^{\dagger}$Correspondence author.}
\thanks{$^{1}$Yubing Cao, Yongming Li, Liejun Wang and Yinfeng Yu are with the School of Computer Science and Technology, Xinjiang University, Urumqi 830049, China (e-mail:{\tt\small 107552201344@stu.xju.edu.cn;
        Lymxju@xju.edu.cn;
        wljxju@xju.edu.cn;
        yuyinfeng@xju.edu.cn;).}}%
\thanks{*This work was supported by these works: the Tianshan Excellence Program Project of Xinjiang Uygur Autonomous Region, China (2022TSYCLJ0036); the Central Government Guides Local Science and Technology Development Fund Projects (ZYYD2022C19); the National Natural Science Foundation of China under Grant 62303259. }
}

\begin{document}

\maketitle
\thispagestyle{empty}
\pagestyle{empty}

\begin{abstract}
Since the introduction of Generative Adversarial Networks (GANs) in speech synthesis, remarkable achievements have been attained. In a thorough exploration of vocoders, it has been discovered that audio waveforms can be generated at speeds exceeding real-time while maintaining high fidelity, achieved through the utilization of GAN-based models. Typically, the inputs to the vocoder consist of band-limited spectral information, which inevitably sacrifices high-frequency details. To address this, we adopt the full-band Mel spectrogram information as input, aiming to provide the vocoder with the most comprehensive information possible.
However, previous studies have revealed that the use of full-band spectral information as input can result in the issue of over-smoothing, compromising the naturalness of the synthesized speech. To tackle this challenge, we propose VNet, a GAN-based neural vocoder network that incorporates full-band spectral information and introduces a Multi-Tier Discriminator (MTD) comprising multiple sub-discriminators to generate high-resolution signals. Additionally, we introduce an asymptotically constrained method that modifies the adversarial loss of the generator and discriminator, enhancing the stability of the training process.
Through rigorous experiments, we demonstrate that the VNet model is capable of generating high-fidelity speech and significantly improving the performance of the vocoder.
\end{abstract}

\section{INTRODUCTION}

Speech synthesis is crucial across various domains, including accessibility, education, entertainment, and customer service \cite{c1}. However, conventional systems often encounter challenges with timbre, speech rate variation, and vocal coherence \cite{c2,c3,c4}. Recent advancements in deep learning and neural network techniques have significantly improved the quality of speech synthesis \cite{c5,c6}. The neural network and deep learning-based speech synthesis now being introduced are broadly divided into two steps: 1) Acoustic modeling: taking characters (text) or phonemes as input and creating a model of the acoustic features. (The acoustic features used in most of the work are Mel Spectrograms); 2) Vocoder: a model that takes Mel Spectrograms (or similar spectrograms) as input and generates real audio \cite{c7}. As an important step in speech synthesis, the study of the vocoder has received extensive attention. This paper focuses on the vocoder part of the study. Vocoder models can be broadly categorized into autoregressive-based (e.g., WaveNet \cite{c8}, WaveRNN \cite{c9}), flow-based (e.g., WaveGlow \cite{c10}, Parallel WaveGAN \cite{c11}), GAN-based \cite{c12} (e.g., MelGAN \cite{c13}, HiFiGAN \cite{c14}, BigVGAN \cite{c15}) and diffusion model-based (e.g., WaveGrad \cite{c16}, Grad-tts \cite{c17}, FastDiff \cite{c18}, ProDiff \cite{c19}) approaches. These advancements promise more natural and coherent speech, enhancing user experience across various applications.

GANs employ an adversarial training approach, where the generator and discriminator engage in a competitive process. This competition fosters improved generator performance and enhances the ability to generate features resembling real data, making GANs widely utilized in vocoder tasks. While the GAN-based generative model can synthesize high-fidelity audio waveforms faster than real-time, most vocoders operate on band-limited Mel spectrogram as input. For instance, HiFi-GAN utilizes band-limited Mel spectrograms as input. Other similar models include LVCNet \cite{c20}, StyleMelGAN \cite{c21} and WaveGlow \cite{c10}. However, speech signals generated with band-limited Mel spectrograms lack high-frequency information, leading to fidelity issues in the resulting waveforms. Thus, considering full-band Mel spectrogram information as vocoder input is crucial. Despite attempts by Parallel WaveGAN to use full-band Mel spectrograms, it faces challenges such as excessive smoothing, resulting in the generation of non-sharp spectrograms and unnatural speech output \cite{c11}.

The loss function of a GAN typically encompasses both the generator and discriminator loss functions. However, various vocoder models employ distinct loss function designs and exhibit differences in the selection of similar loss terms, leading to training instability. For instance, Parallel WaveGAN incorporates cross-entropy loss into the generator loss to address instability issues, albeit without complete resolution \cite{c11}. MelGAN endeavors to enhance stability by replacing the cross-entropy loss with hinge loss and augmenting feature matching loss, yet gradient loss persists \cite{c13}. HiFiGAN introduces feature matching loss and Mel spectrogram loss to mitigate training instability \cite{c14}. Despite the inclusion of these additional loss functions, training may still encounter challenges such as gradient loss and pattern collapse, resulting in an unstable training process.

\begin{figure*}[t]
\begin{minipage}[b]{1.0\linewidth}
  \centering
  \centerline{\includegraphics[width=18cm,height=9cm]{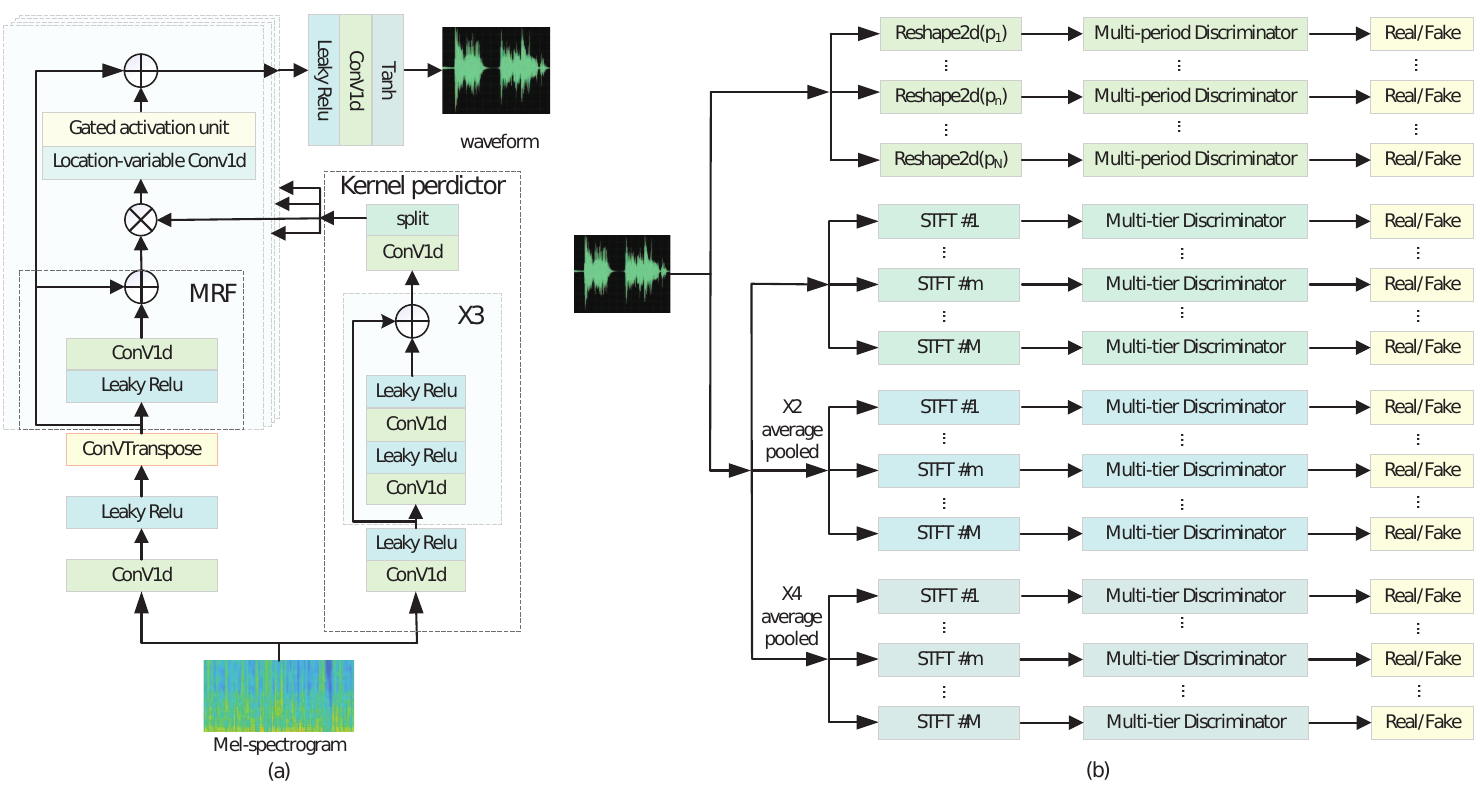}}
  \caption{VNet structure. (a) is the structure of the generator, and (b) is the structure of the discriminator. “STFT\#m” means the process of calculating the amplitude of a linear spectrogram using the m-th STFT parameter set, and “reshape2d(p)” means the process of reshaping a 1D signal of length T into a 2D signal of height T/p and width p.}\medskip
  \label{fig:1}
\end{minipage}
\end{figure*}

This paper introduces VNet, a novel vocoder model capable of synthesizing high-fidelity speech in real time. A new discriminator module, named MTD, is proposed, which utilizes multiple linear spectrogram magnitudes computed with distinct sets of parameters. Operating on full-band Mel spectrogram data, MTD facilitates the generation of full-band and high-resolution signals. The overall discriminator integrates a Multi-Period Discriminator (MPD), leveraging multiple scales of waveforms to enhance speech synthesis performance by capturing both time and frequency domain characteristics \cite{c14}. To mitigate model training instability, an asymptotically constrained approach is proposed to modify the adversarial training loss function. This entails constraining the adversarial training loss within a defined range, ensuring stable training of the entire model. Our contributions can be summarized in three main aspects:
\begin{itemize}

\item We propose VNet, a neural vocoder network for GAN-based speech synthesis that incorporates an MTD module to capture the features of speech signals from both time and frequency domains.
\item We propose an asymptotically constrained approach to modify the adversarial training loss of the generator and discriminator of the vocoder.
\item We demonstrate the effectiveness of the VNet model, as well as the effectiveness of the newly added MTD module and asymptotic constraints against training loss.

\end{itemize}

\section{RELATED WORK}

GANs have emerged as powerful generative models \cite{c12}. Initially applied to image generation tasks, GANs have garnered significant success and attention. Similarly, in the domain of speech synthesis, where traditional approaches primarily rely on rule-based or statistical models, GAN technology has gradually gained traction. By leveraging the adversarial framework of GANs, speech synthesis models can better capture the complexity and realism of speech signals, thereby producing more natural, high-quality synthesized speech. 

WaveGAN simplifies speech synthesis by directly generating raw audio waveforms, producing high-quality and naturalistic speech segments. However, its training requires substantial data and computational resources. In contrast, Parallel WaveGAN extends single short-time Fourier transform (STFT) loss to multi-resolution, integrating it as an auxiliary loss for GAN training\cite{c11}. It may suffer from excessive smoothing. MelGAN achieves high-quality synthesis without additional distortion or perceptual losses by introducing a multi-scale discriminator (MSD) and incorporating hinge loss, feature matching loss, and discriminator loss\cite{c13}. HiFiGAN enhances the discriminator's ability to differentiate between generated and real audio and introduces a multi-receptive field fusion (MRF) module in the generator. Its loss functions include least squares loss, feature matching loss, Mel spectrogram loss, and discriminator loss\cite{c14,c22}. BigVGAN builds upon HiFiGAN by replacing the MSD with a multi-resolution discriminator (MRD) and introducing periodic activation into the generator. It proposes an anti-aliasing multi-periodicity composition (AMP) module for modeling complex audio waveforms. BigVGAN's loss functions comprise least squares adversarial loss, feature matching loss, and Mel spectrogram loss\cite{c15}.VNet distinguishes itself from these methods by simultaneously addressing the challenges of matching features at various resolutions and scales while also resolving the issue of poor fidelity results that arise from using full-band Mel spectrograms as input.

\section{METHOD}

\subsection{Generator} The generator G, inspired by BigVGAN, is a fully convolutional neural network depicted in Fig. \ref{fig:1}(a). It takes a full-band Mel spectrogram as input and utilizes inverse convolution for upsampling until the output sequence length matches the target waveform map. Each deconvolution module is followed by an MRF module, which concurrently observes pattern features of varying lengths. The MRF module aggregates the outputs of multiple residual modules, each with different convolution kernel sizes and expansion coefficients, aimed at forming diverse perceptual field patterns.

To efficiently capture localized information from the Mel spectrogram, we introduce Location Variable Convolution (LVC), enhancing sound quality and generation speed while maintaining model size \cite{c20}. The LVC layer's convolution kernel is obtained from the kernel predictor, with the Mel spectrogram serving as input and the predicted convolution kernel concatenated into a residual stack for each LVC layer separately. Through empirical experiments, we optimize the placement and number of LVC layers and the kernel predictor to achieve the desired sound quality and generation speed. To improve the model's adaptability to speaker feature variations and mitigate overfitting risks, we incorporate gated activation units (GAUs) \cite{c23}.

\begin{figure}[t]
\begin{minipage}[b]{1.0\linewidth}
  \centering
  \centerline{\includegraphics[width=9cm,height=9cm]{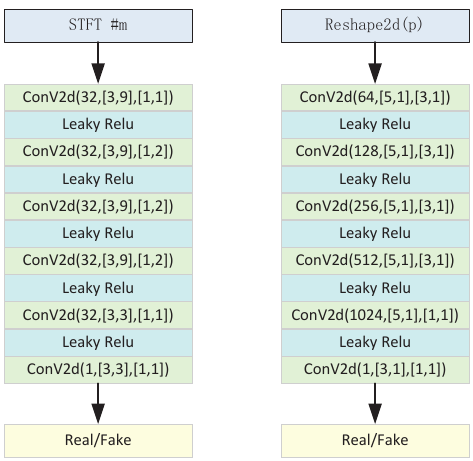}}
  \caption{Details of the discriminator, left: MTD, where the values in parentheses denote (output channel, kernel width, expansion rate), respectively. Right: MPD, where the values in parentheses denote (output channel, [kernel width, kernel height], [step width, step height]), respectively.}\medskip
  \label{fig:2}
\end{minipage}
\end{figure}

\begin{figure*}[t]
\begin{minipage}[b]{0.3\linewidth}
  \centering
  \centerline{\includegraphics[width=6cm,height=5cm]{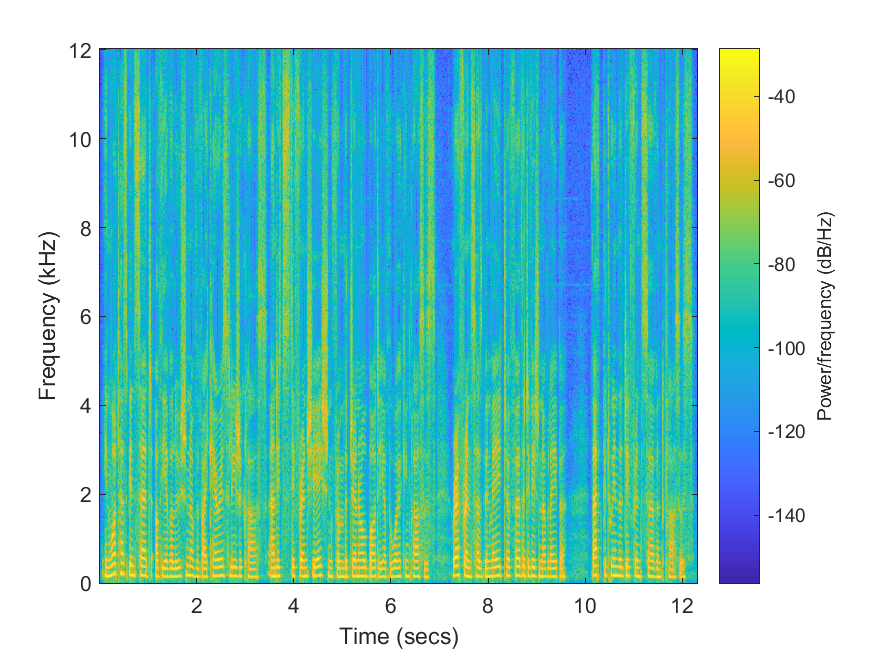}}
  \centerline{(a) Ground Truth}\medskip
  \label{subfig:3(a)}
\end{minipage}
\hfill
\begin{minipage}[b]{0.3\linewidth}
\centering
  \centerline{\includegraphics[width=6cm,height=5cm]{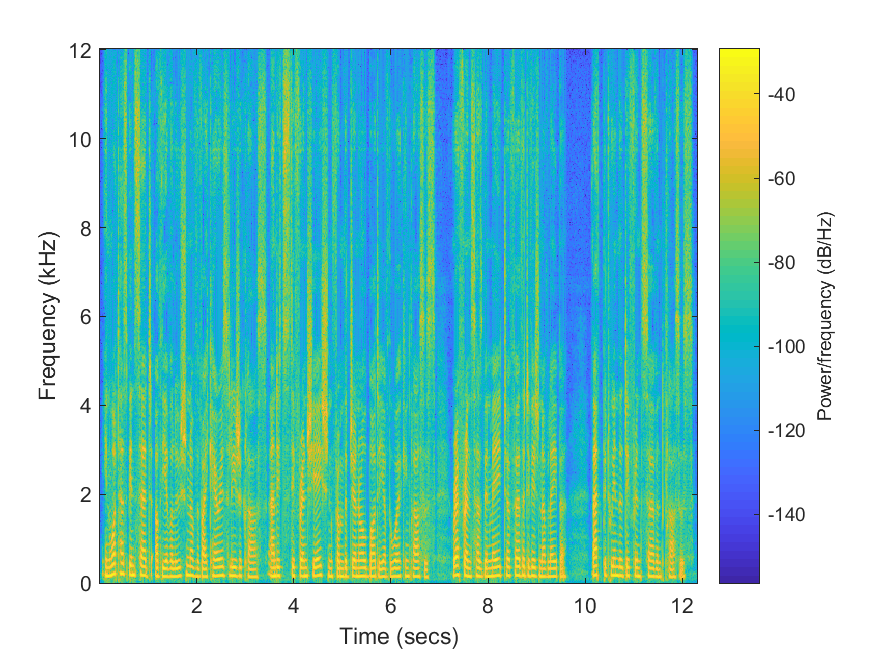}}
  \centerline{(b) BigVGAN}\medskip
  \label{subfig:3(b)}
\end{minipage}
\hfill
\begin{minipage}[b]{0.3\linewidth}
\centering
  \centerline{\includegraphics[width=6cm,height=5cm]{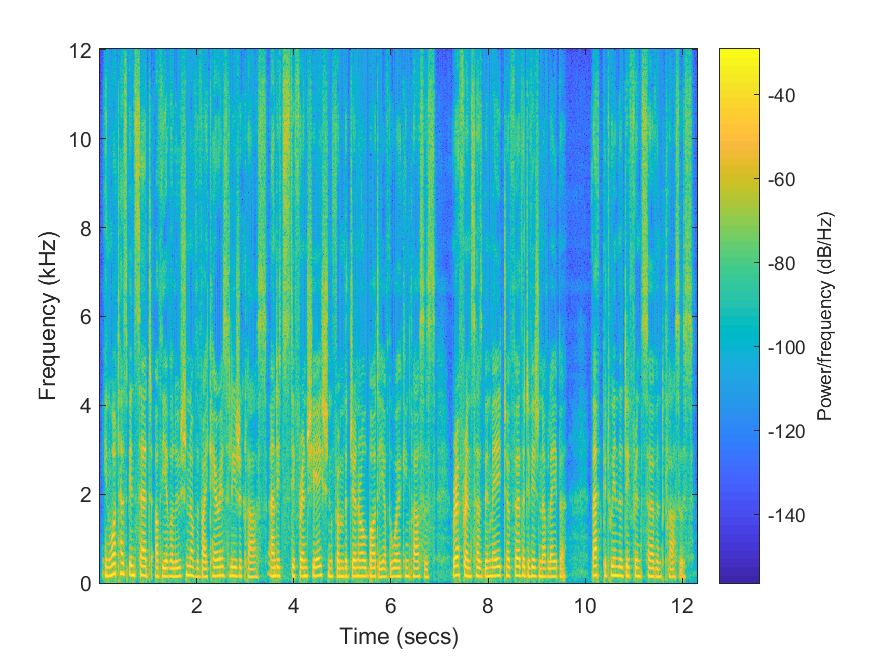}}
  \centerline{(c) VNet}\medskip
  \label{subfig:3(c)}
\end{minipage}
\caption{Spectrograms of synthesized samples with BigVGAN and VNet trained on the LibriTTS train set for 1M steps and the corresponding ground truth.}\medskip
\label{fig:3}
\end{figure*}

\subsection{Discriminator} Discriminators play a crucial role in guiding the generator to produce high-quality, coherent waveforms while minimizing perceptual errors detectable by the human ear. State-of-the-art GAN-based vocoders typically incorporate multiple discriminators to guide coherent waveform generation while minimizing perceptual artifacts. Moreover, each discriminator comprises several sub-discriminators. As illustrated in Fig. \ref{fig:1}(b), our discriminator utilizes multiple spectrograms and reshaped waveforms computed from real or generated signals. Since speech signals contain sinusoidal signals with varying periods, we introduce the MPD to identify various periodic patterns in the audio data. MPD extracts periodic components from waveforms at prime intervals and utilizes them as inputs to each subsampler \cite{c14}. Additionally, to capture continuous patterns and long-term dependencies, we design and employ the MTD.

MTD comprises three sub-discriminators operating at different input scales: raw audio, ×2 average pooled audio, and ×4 average pooled audio. Each sub-discriminator receives input from the same waveform through STFT using distinct parameter sets \cite{c11}. These parameter sets specify the number of points in the Fourier transform, frame-shift interval, and window length.

Each sub-discriminator in MTD consists of stride and packetized convolutional layers with Leaky ReLU activation. The mesh size increases by reducing the step size and adding more layers. Spectral normalization stabilizes the training process, except for the first subframe, where weight normalization manipulates raw audio. This model architecture draws inspiration from Multi-Scale Waveform Diagrams (MSWDs) but diverges by utilizing MTD to incorporate multiple spectrograms with varying temporal and spectral resolutions, thereby generating high-resolution signals across the full frequency band.

The VNet discriminator comprises two sub-modules: the MTD and the MPD, each containing multiple sub-discriminators utilizing 2D convolutional stacking, as depicted in Fig. \ref{fig:2}, MTD transforms the input 1D waveform into a 2D linear spectrogram by employing various downsampling average pooling multiples, followed by STFT with diverse parameters ([n\_fft, hop\_length, win\_length]). MPD converts the input 1D waveform of length T into a 2D waveform through reshaping and reflection filling (Reshape2d) with different widths (p) and heights (T /p).

\subsection{Training Losses}

The feature matching loss measures similarity in learning, quantifying the difference in sample features between ground truth and generated samples \cite{c24}. Given its successful application in speech synthesis, we employed it as an additional loss for training the generator. Each intermediate feature was extracted, and the Frobenius distance between ground truth and generated samples in each feature space was computed. Denoted as $L_{FM}$, the feature matching loss is defined as follows:
$$
L_{FM}(X,\hat{X})=\frac{1}{M}\sum_{m=1}^{M}E_{X,\hat{X}}[\frac{||S_m-\hat{S_m}||_F}{||S_m||_F}]\quad, \eqno{(1)}
$$
where$||\cdot||_F$denote the Frobenius norms and S denotes the number of elements in the spectrogram. Each m-th L$_{FM}$ reuse $S_m$ and $\hat{S_m}$ used in the m-th MTD sub-discriminator. The number of each loss is M, which is the same as the number of MTD sub-discriminators.

We also introduced a log-Mel spectrogram loss to enhance the training efficiency of the generator and improve the fidelity of the generated audio. Drawing from previous work, incorporating reconstruction loss into the GAN model has been shown to yield realistic results \cite{c25}. We employed the Mel spectrogram loss based on input conditions, aiming to focus on improving perceptual quality given the characteristics of the human auditory system \cite{c26}. The Mel spectrogram loss is calculated as the L1 distance between the Mel spectrogram of the waveforms generated by the generator and the Mel spectrogram of the ground truth waveforms. Denoted as L$_{Mel}$, the Mel spectrogram loss is defined as follows:
$$
L_{Mel}(X,\hat{X})=\frac{1}{M}\sum_{m=1}^{M}E_{X,\hat{X}}[\frac{1}{S_m}]||logS_m-log\hat{S_m}||_1\quad, \eqno{(2)}
$$
where$||\cdot||_1$ denotes the L1 norms, and S denotes the number of elements in the spectrogram. Each m-th L$_{FM}$ reuse  $S_m$ and $\hat{S_m}$ used in the m-th MTD sub-discriminator. The number of each loss is M, which is the same as the number of MTD sub-discriminators.

 The objective of the vocoder is to train the generating function$G_\theta S\rightarrow X$, which transforms a Mel spectrogram $s\in S $ into a waveform signal $x\in X $. The adversarial losses of the generator and the discriminator are denoted as $L_{adv}(G; D)$ and $L_{adv}(D; G)$. The discriminant function $D: X \in R $ is typically implemented using a neural network, denoted by $\phi$, which comprises linear operations and nonlinear activation functions \cite{c27}. To simplify, we decompose the discriminator into a nonlinear function $h_\varphi:X \in W\subseteq R^D$ and a final linear layer $\omega \in W$, expressed as $D^W_\varphi (x)=W^Th_\varphi (x)$, where $\phi =[{\varphi,\omega}]$. The discriminative process can be interpreted as segmenting the nonlinear feature $h_\varphi (x)$ using a shared projection $\omega$. Thus, the adversarial loss of the generator and the discriminator can be expressed as follows:
$$
L_{adv}(D;G)=E_{PX}[R_1(D_\varphi^\omega(X))]+E_{PS}[R_2(D_\varphi^\omega(G_\theta(s))], \eqno{(3)}
\label{Eq: 3}
$$

$$
L_{adv}(G;D)=E_{PS}[R_3(D_\varphi^\omega(G_\theta(s)))], \eqno{(4)}
$$

$$
R_1(z)=-(1-z)^2, R_2(z)=-z^2, R_3(z)=(1-z)^2, \eqno{(5)}
$$
$p_X (x)$ and $p_S (s)$ denote the waveform signal and Mel spectrogram, respectively. Through optimization of the maximization problem, a nonlinear function $h_\varphi$ is induced to differentiate between true and false samples and mapped onto the feature space W, resulting in a linear projection on W to enhance discrimination \cite{c28}. However, the linear projection $\omega$ in Eq. (3) may not fully utilize features for discrimination. We observe that given $h_\varphi$, there exist linear projections that offer more discriminative information than the projection $\omega$ maximizing Eq. (3). As long as $R_3$ (whose derivative is denoted $r_3$ ) is a monotonically decreasing function—meaning the derivative $r_3 (z)$ is negative for any $z \in R$. Thus, we propose the asymptotic constraint method to modify the adversarial loss function of the generator and the discriminator as follows:
\begin{table*}[t]
\caption{objective and subjective evaluations on libritts. objective results are obtained from a subset of its dev set. subjective evaluations are based on a 5-scale mean opinion score (mos) with 95\% confidence interval (ci) from a subset of its test set. “speed” indicates how fast each model is generated relative to real-time.}
\label{table1}
\begin{center}
\begin{tabular}{|c|c|c|c|c|c|c|c|c|}
\hline
LibriTTS & M-STFT($\downarrow$) & PESQ($\uparrow$) & MCD($\downarrow$) & Periodicity($\downarrow$) & V/UF $F_1$($\uparrow$) & MOS($\uparrow$) & Paraments & Speed\\
\hline
Parrel WaveGAN & 1.3422 & 3.642 & 2.1548 & 0.1478 & 0.9359 & 3.88$\pm$0.09 & 4.34M & 294.11$\times$\\
\hline
Wave Glow & 1.3099 & 3.138 & 2.3591 & 0.1485 & 0.9378 & 3.84$\pm$0.10 & 99.43M & 31.87$\times$\\
\hline
HiFi-GAN & 1.0017 & 2.947 & 0.6603 & 0.1565 & 0.9300 & 4.08$\pm$0.09 & 14.01M & 135.14$\times$\\
\hline
BigVGAN & 0.7997 & 4.027 & 0.3745 & 0.1018 & 0.9598 & 4.11$\pm$0.09 & 112.4M & 44.72$\times$\\
\hline
VNet(Ours) & \textbf{0.7892} & \textbf{4.032} & \textbf{0.3711} & \textbf{0.0948} & \textbf{0.9637} & 4.13$\pm$0.09 & 14.86M & 204.08$\times$\\
\hline
Ground Truth & - & - & - & - & - & \textbf{4.40$\pm$0.06} & - & -\\
\hline
\end{tabular}
\end{center}
\end{table*}

\begin{align}
    L_{adv}(D;G) 
    & = E_{PX}[R_1(D_\varphi^{\omega^-}(X))] + E_{PS}[R_2(D_\varphi^{\omega^-}(G_\theta(s)))] \nonumber \\
    & \quad + E_{PX}[R_3(D_{\varphi^-}^\omega(X))] - E_{PS}[R_3(D_{\varphi^-}^\omega(G_\theta(s)))] \tag{6}, 
    \label{Eq: 6}
\end{align}
$$
L_{adv}(G;D)=E_{PS}[R_3(D_\varphi^\omega(G_\theta(s)))], \eqno{(7)}
$$

$$
R_1(z)=-\sigma(1-z)^2, R_2(z)=-\sigma(z)^2, R_3(z)=\sigma(1-z)^2, \eqno{(8)}
$$
where $\sigma(\cdot)$ is the "asymptotic constraint", i.e., $\sigma(x)=e^{-(0.3x-2)}$. In our preliminary experiments, when utilizing Eq. (3) instead of Eq. (6), we observed unstable training, underscoring the significance of ensuring the monotonicity of $\hat{R_3}$. Particularly in the early stages of training, the loss values tended to converge to suboptimal local minima.

\section{EXPERIMENTS}

\subsection{Data configurations}

We validate the effectiveness of our method on the LibriTTS dataset, an English multi-speaker audiobook dataset comprising 585 hours of audio \cite{c29}. The training utilizes the LibriTTS training sets (train-clean-100, train-clean-360 and train-other-500). For text-to-speech evaluation, we fine-tune the vocabulary encoder using predicted log-mel spectrograms to minimize feature mismatches. Additionally, we employ the LJSpeech dataset\footnote{https://keithito.com/LJ-Speech-Dataset}, containing 24 hours of data and 13,000 utterances from English-speaking female speakers. All waveforms are sampled at a rate of 24 kHz.

All models, including the baseline, are trained using a frequency range of [0, 12] kHz and 100-band logarithmic Mel spectrograms, consistent with recent studies on universal vocoders. STFT parameters are set as per previous work, with a 1024 FFT size, a 1024 Hann window, and a 256 hop size. Objective evaluation is conducted on a subset of LibriTTS dev-clean and dev-other. Following the formal implementation of VNet, evaluation involves 6\% randomly selected audio files from dev-clean and 8\% randomly selected audio files from dev-other. For this experiment, we utilized
a server with 4 Tesla T4 GPUs, each with a 16GB memory capacity. The CPU
used in the server is an Intel Xeon Gold 5218R.

\subsection{Evaluation metrics}

We conduct an objective assessment using five metrics: 1) multi-resolution STFT (M-STFT)\footnote{https://github.com/ludlows/PESQ} measuring the spectral distance between the multiple resolutions; 2) perceptual evaluation of speech quality (PESQ)\footnote{https://github.com/ttslr/python-MCD}\, a widely adopted method for automated speech quality assessment \cite{c30}; 3) mel-cepstral distortion (MCD), quantifying differences between resolutions using dynamic time warping \cite{c31}; 4) periodicity error and 5) F1 scores for voiced/unvoiced classification (V/UV F1)\footnote{https://github.com/descriptinc/cargan}, capturing main artifacts from non-autoregressive GAN-based vocoders \cite{c32}. Metrics are computed on each subset and then macro-averaged across subsets.

Additionally, Mean Opinion Score (MOS) tests are conducted on a combination of Test Clean and Test Other sets. Eight raters evaluate the synthesized speech samples using a five-point scale: 1 = Bad; 2 = Poor; 3 = Fair; 4 = Good; 5 = Excellent. Ten utterances are randomly selected from the combined test set and synthesized using the trained model. It's important to note that MOS is a relative metric, with listeners utilizing the entire scale regardless of the absolute quality of the samples in the test.

\subsection{Comparison with existing models}

Table \ref{table1} presents the results, with Wave Glow and Parallel WaveGAN yielding lower scores than other models and VNet outperforming BigVGAN across all objective and subjective evaluations. While there's only a marginal enhancement in subjective scores compared to HiFi-GAN, VNet offers the advantage of generating results at approximately 1.5 times the speed for a similar number of parameters. Notably, Parallel WaveGAN exhibits over-smoothing issues, likely due to experiments with full-band features instead of band-limited features.

As depicted in Fig. \ref{fig:3}, PESQ only considers the [0, 8] kHz range, while MCD and M-STFT assess both this range and higher frequency bands, resulting in significantly improved MCD and M-STFT scores. MOS scores demonstrate a strong correlation with PESQ scores.

\subsection{Ablation study}

In order to further validate the significance of each component in our proposed model VNet, we conducted qualitative and quantitative analyses on the speech generated by the generator. We systematically removed specific key architectural components and evaluated the audio quality using a designated test set. Table \ref{table2} presents the average opinion scores of the audio quality assessed through human listening tests. Each model underwent training on the LJSpeech dataset for 400k iterations.
Our analysis indicates that solely utilizing MPD without incorporating other discriminators leads to skipping certain segments of the sound, resulting in the loss of some words in the synthesized speech. Incorporating MSD alongside MPD 
\begin{table}[h]
\caption{results of ablation experiments on the ljspeech dataset}
\label{table2}
\begin{center}
\begin{tabular}{|>{\centering\arraybackslash}m{3cm}|>{\centering\arraybackslash}m{2cm}|}
\hline
Model & MOS($\uparrow$)\\
\hline
w/o MTD & 3.74$\pm$0.09\\
\hline
w MPD $\delta$ MRD & 3.25$\pm$0.09\\
\hline
w MPD $\delta$ MSD & 3.35$\pm$0.09\\
\hline
w/o Modified $L_{adv}$ & 3.08$\pm$0.09\\
\hline
VNet(Ours) & \textbf{4.13$\pm$0.09}\\
\hline
\end{tabular}
\end{center}
\end{table}
improves the retention of words yet makes it challenging to capture sharp high-frequency patterns, resulting in samples that sound noisy. The addition of MRD to MPD further enhances word retention but introduces metallic artefacts in the audio, which are particularly noticeable during speaker breathing intervals.

\section{CONCLUSIONS AND FUTURE WORK}

This study demonstrates the capabilities of the VNet model, a GAN-based vocoder, in enhancing speech synthesis. By utilizing full-band Mel spectrogram inputs, the model effectively addresses over-smoothing issues. Furthermore, the introduction of a Multi-Tier Discriminator (MTD) and refined adversarial loss functions has significantly improved speech quality and fidelity.

Future research should prioritize further reducing over-smoothing and exploring the model's potential in multilingual and diverse speech styles. Such advancements could greatly enhance the practical usability of GAN-based vocoders, resulting in more natural and expressive synthesized speech.

\addtolength{\textheight}{-12cm}   






\begin{thebibliography}{99}

\bibitem{c1} D. D. Lim, W. Jang, G. O, H. Park, B. Kim, and J. Yoon, "JDI-T: Jointly trained Duration Informed Transformer for Text-To-Speech without Explicit Alignment," in \textit{Proceedings of the Annual Conference of the International Speech Communication Association (INTERSPEECH)}, 2020.
\bibitem{c2} M. Morise, F. Yokomori, and K. Ozawa, "WORLD: A vocoder-based high-quality speech synthesis system for real-time applications," \textit{IEICE Transactions on Information and Systems}, vol. 99, no. 7, pp. 1877-1884, 2016.
\bibitem{c3} D. Griffin and J. Lim, "Signal estimation from modified short-time Fourier transform," IEEE Transactions on Acoustics, Speech, and Signal Processing, vol. 32, no. 2, pp. 236-243, 1984.
\bibitem{c4} H. Kawahara, I. Masuda-Katsuse, and A. De Cheveigne, "Restructuring speech representations using a pitch-adaptive time--frequency smoothing and an instantaneous-frequency-based F0 extraction: Possible role of a repetitive structure in sounds," Speech Communication, vol. 27, no. 3-4, pp. 187-207, 1999.
\bibitem{c5} Y. Yu, Z. Jia, F. Shi, M. Zhu, W. Wang, and X. Li, "WeaveNet: End-to-End Audiovisual Sentiment Analysis," in \textit{International Conference on Cognitive Systems and Signal Processing}, Springer, 2021, pp. 3-16. 
\bibitem{c6} Y. Rn, C. Hu, X. Tan, T. Qin, S. Zhao, Z. Zhao, and T.-Y. Liu, "FastSpeech 2: Fast and high-quality end-to-end text to speech," in \textit{Proc. ICLR}, 2021.
\bibitem{c7} T. Shibuya, Y. Takida, and Y. Mitsufuji, "Bigvsan: Enhancing GAN-based neural vocoders with slicing adversarial network," in \textit{ICASSP 2024-2024 IEEE International Conference on Acoustics, Speech and Signal Processing (ICASSP)}, IEEE, 2024, pp. 10121-10125.
\bibitem{c8} A. Van Den Oord, S. Dieleman, H. Zen, et al., "Wavenet: A generative model for raw audio," \textit{arXiv preprint arXiv:1609.03499}, Dec. 2016.
\bibitem{c9} N. N. Kalchbrenner et al., "Efficient neural audio synthesis," in \textit{International Conference on Machine Learning}, PMLR, 2018, pp. 2410-2419. 
\bibitem{c10} R. R. Prenger, R. Valle, and B. Catanzaro, "Waveglow: A flow-based generative network for speech synthesis," in \textit{ICASSP 2019-2019 IEEE International Conference on Acoustics, Speech and Signal Processing (ICASSP)}, IEEE, 2019, pp. 3617-3621.
\bibitem{c11} S. S. Yamamoto, E. Song, and J.-M. Kim, "Parallel WaveGAN: A fast waveform generation model based on generative adversarial networks with multi-resolution spectrogram," in \textit{ICASSP 2020-2020 IEEE International Conference on Acoustics, Speech and Signal Processing (ICASSP)}, IEEE, 2020, pp. 6199-6203.
\bibitem{c12} I. Goodfellow, J. Pouget-Abadie, M. Mirza, et al., "Generative adversarial nets," \textit{Advances in Neural Information Processing Systems}, vol. 27, 2014.
\bibitem{c13} K. Kumar, R. Kumar, T. De Boissiere, et al., "Melgan: Generative adversarial networks for conditional waveform synthesis," \textit{Advances in Neural Information Processing Systems}, vol. 32, 2019.
\bibitem{c14} J. J. Kong, J. Kim, J. Bae, "Hifi-gan: Generative adversarial networks for efficient and high fidelity speech synthesis," \textit{Advances in Neural Information Processing Systems}, vol. 33, pp. 17022-17033, 2020.
\bibitem{c15} S. -g. Lee, W. Ping, B. Ginsburg, B. Catanzaro, and S. Yoon, "Bigvgan: A universal neural vocoder with large-scale training," \textit{arXiv preprint arXiv:2206.04658}, 2022.
\bibitem{c16} O. O.Chen, Y. Zhang, H. Zen, R. J. Weiss, M. Norouzi, and W. Chan, "Wavegrad: Estimating gradients for waveform generation," \textit{arXiv preprint arXiv:2009.00713}, 2020.
\bibitem{c17} V. Popov, I. Vovk, V. Gogoryan, T. Sadekova, and M. Kudinov, "Grad-tts: A diffusion probabilistic model for text-to-speech," in International Conference on Machine Learning, 2021, pp. 8599-8608, PMLR.
\bibitem{c18}  R. R.Huang, M. W. Y. Lam, J. Wang, D. Su, D. Yu, Y. Ren, and Z. Zhao, "Fastdiff: A fast conditional diffusion model for high-quality speech synthesis," \textit{arXiv preprint arXiv:2204.09934}, 2022.
\bibitem{c19} S. S. Huang, Z. Zhao, H. Liu, J. Liu, C. Cui, and Y. Ren, "Prodiff: Progressive fast diffusion model for high-quality text-to-speech," in \textit{Proceedings of the 30th ACM International Conference on Multimedia}, 2022, pp. 2595-2605.
\bibitem{c20} Z. Z. Zeng, J. Wang, N. Cheng, and J. Xiao, "Lvcnet: Efficient condition-dependent modeling network for waveform generation," in \textit{ICASSP 2021-2021 IEEE International Conference on Acoustics, Speech and Signal Processing (ICASSP)}, IEEE, 2021, pp. 6054-6058. 
\bibitem{c21} A. A. Mustafa, N. Pia, and G. Fuchs, "Stylemelgan: An efficient high-fidelity adversarial vocoder with temporal adaptive normalization," in \textit{ICASSP 2021-2021 IEEE International Conference on Acoustics, Speech and Signal Processing (ICASSP)}, IEEE, 2021, pp. 6034-6038. 
\bibitem{c22} X. X. Mao et al., "Least squares generative adversarial networks," in \textit{Proceedings of the IEEE International Conference on Computer Vision}, 2017, pp. 2794-2802.
\bibitem{c23} A. A. Van den Oord et al., "Conditional image generation with PixelCNN decoders," \textit{Advances in Neural Information Processing Systems}, vol. 29, 2016.
\bibitem{c24} Z. Guo, G. Yang, D. Zhang, and M. Xia, "Rethinking gradient operator for exposing AI-enabled face forgeries," \textit{Expert Systems with Applications}, vol. 215, pp. 119361, 2023.
\bibitem{c25} P. P. Isola, J.-Y. Zhu, T. Zhou, and A. A. Efros, "Image-to-image translation with conditional adversarial networks," in \textit{Proceedings of the IEEE Conference on Computer Vision and Pattern Recognition}, 2017, pp. 1125-1134. 
\bibitem{c26} W. W. Ping, K. Peng, and J. Chen, "Clarinet: Parallel wave generation in end-to-end text-to-speech," \textit{arXiv preprint arXiv:1807.07281}, 2018.
\bibitem{c27} X. Jiao, L. Wang, and Y. Yu, "MFHCA: Enhancing Speech Emotion Recognition Via Multi-Spatial Fusion and Hierarchical Cooperative Attention," \textit{arXiv preprint arXiv:2404.13509}, 2024. 
\bibitem{c28} C. Bollepalli, L. Juvela, and P. Alku, "Generative adversarial network-based glottal waveform model for statistical parametric speech synthesis," in \textit{Proc. Interspeech}, 2017, pp. 3394–3398.
\bibitem{c29} H. H. Zen, V. Dang, R. Clark, Y. Zhang, R. J. Weiss, Y. Jia, Z. Chen, and Y. Wu, "LibriTTS: A corpus derived from LibriSpeech for text-to-speech," \textit{arXiv preprint arXiv:1904.02882}, 2019.
\bibitem{c30} A. A. W. Rix, J. G. Beerends, M. P. Hollier, and A. P. Hekstra, "Perceptual evaluation of speech quality (PESQ)-a new method for speech quality assessment of telephone networks and codecs," in \textit{2001 IEEE International Conference on Acoustics, Speech, and Signal Processing. Proceedings (Cat. No. 01CH37221)}, vol. 2, pp. 749-752, 2001.
\bibitem{c31} R. R. Kubichek, "Mel-cepstral distance measure for objective speech quality assessment," in \textit{Proceedings of IEEE Pacific Rim Conference on Communications Computers and Signal Processing}, vol. 1, 1993, pp. 125-128.
\bibitem{c32} L. L. Morrison, R. Kumar, K. Kumar, P. Seetharaman, A. Courville, and Y. Bengio, "Chunked autoregressive GAN for conditional waveform synthesis," in \textit{Proc. ICLR}, 2022.


\end{thebibliography}
\end{document}